\documentclass[a4paper,11pt]{article}
\pdfoutput=1 

\usepackage{jcappub} 

\usepackage[T1]{fontenc} 
\usepackage[normalem]{ulem}

\newcommand{\bea}{\begin{eqnarray}}
\newcommand{\eea}{\end{eqnarray}}

\newcommand{\beq}{\begin{equation}}
\newcommand{\eeq}{\end{equation}}

\newcommand{\simless}[0]{\mathbin{\lower 3pt\hbox
   {$\rlap{\raise 5pt\hbox{$\char'074$}}\mathchar"7218$}}}
\newcommand{\simgreat}[0]{\mathbin{\lower 3pt\hbox
   {$\rlap{\raise 5pt\hbox{$\char'076$}}\mathchar"7218$}}}

\newcommand{\figref}[1]{figure \eqref{#1}}

\newcommand{\capfigref}[1]{Figure \eqref{#1}}

\newcommand{\eqnref}[1]{eq.\hspace{1mm}\eqref{#1}} 
\newcommand{\eqnrefs}[1]{eqs. \eqref{#1}} 
\newcommand{\eqnrefbare}[1]{\eqref{#1}} 

\title{Non-linear evolution in $f(R)$ gravity: perturbative modelling of the Chameleon mechanism  }

\author[a,1]{Sharvari Nadkarni-Ghosh \note{Corresponding author.}}
\author[b]{and Tanush Reddy Vaka}

\affiliation[a]{Department of SPASE, Indian Institute of Technology (I.I.T.) Kanpur,\\ Kanpur, U.P. 208016 India }
\affiliation[b]{School of Physical Sciences, National Institute of Science, Education and Research (NISER) \\ Jatni, Khorda, Odisha 752050 India }

\emailAdd{sharvari@iitk.ac.in}
\emailAdd{tanushreddy.vaka@niser.ac.in}

\abstract{We investigate the non-linear evolution of matter perturbations in $f(R)$ models with the Chameleon screening mechanism. The novel feature of our investigation is an iterative solution for the non-linear equation for the scalar field $\chi = \Phi - \Psi$, where $\Phi$ and $\Psi$ are the potentials that characterise scalar perturbations of the metric. We demonstrate the scheme on spherical perturbations - smooth, compensated top-hats of varying length scales. We find that the effect of the Chameleon mechanism is seen most prominently on scales where the size of the top-hat is comparable to the Compton scale of the background. There is a density enhancement near the outer edge of the top-hat and the top-hat does not retain its shape. We explain this well-known observation in the context of the spatio-temporal evolution of the Compton scale. Additionally, we find a slight enhancement of the density near the origin, a feature not reported previously in the literature. On scales much smaller or much larger than the background Compton length, including the Chameleon screening has  no appreciable effect on the perturbations. In the former, the growth is enhanced as compared to GR and is almost the same as GR in the latter. Finally, we examine the non-linear density  
velocity divergence (DVDR) relation and find that for evolution affected by Chameleon screening, the DVDR is no longer one-to-one even for a single profile. The relation between density and velocity depends on the location within the perturbation. }

\begin{document}
\maketitle
\flushbottom

\section{Introduction}
\label{sec:intro}
The standard $\Lambda$CDM model of cosmology faces multiple challenges. On the theoretical front, the origins of dark matter and dark energy are unknown. 
On the observational front, there are several inconsistencies in the data, most notably the Hubble and $\sigma_8$ tensions  (for e.g., ~\cite{Macaulay_2013,Di_Valentino_2021}). Thus, there is a pressing need to investigate theories beyond the standard model. Broadly speaking, there are two approaches to modifying the standard model. Either the nature of dark energy is modified from the theoretically proposed cosmological constant to some other phenomenological model such as quintessence ~\citep{Peebles_2003,Tsujikawa_2013,Tsujikawa_2015}, or the Einsteinian action is modified, which gives rise to equations that are different from those of the standard GR model (for e.g., \cite{sotiriou_fr_2010,defelice_fR_2010,clifton_modified_2012,Nojiri_2017}). For some class of models, these two approaches are rendered equivalent by certain mathematical transformations but more generally, they are different (see for e.g., \cite{Capozziello_2006,Capozziello_2006b, joyce_beyond_2015, joyce_dark_2016, wettrich_2015,Bahamonde_2017,odinstov2025}). In most modifications, the additional degree of freedom introduced by the modification allows one to recover the background $\Lambda$CDM expansion history exactly. Therefore, observations that rely on the background geometry of the Universe, alone, are not enough to distinguish these models from each other or from $\Lambda$CDM. Thus, modifications to $\Lambda$CDM necessarily need to be constrained by tracking the growth of perturbations. Indeed, many current and upcoming surveys such as  SDSS\footnote{https://www.sdss.org},  Euclid\footnote{http://sci.esa.int/euclid/.}, DESI\footnote{https://www.desi.lbl.gov/} and the Vera C. Rubin Observatory \footnote{https://www.lsst.org} all aim to measure the matter distribution at different epochs with the primary aim of constraining cosmological parameters \citep{alam_SDSS_2017,amendola_2018,alam_testing_2020,pal_2025,pal_2026}.

Einstein's gravity has been tested very well on solar system scales using various effects such as perihelion precession, gravitational redshift  etc. (see \citep{Turyshev_2008,Will_2014} for comprehensive reviews). Thus, any viable modification has to include a screening mechanism through which Einstein's gravity can be recovered in high density, strong gravity regimes such as the solar system, while allowing for deviations on larger cosmological scales and lower density contrasts. In this paper, we focus on one particular class of models, the $f(R)$ models, wherein the Ricci scalar $R$ in the Einstein-Hilbert action is replaced with a general function $f(R)$. The screening mechanism in these models is referred to as the `Chameleon Mechanism' \citep{khoury_chameleon_2004}. In this mechanism, the Compton scale associated with the extra degree of freedom is density-dependent. In high density regions, the Compton scale drops to a very small value and effects of the modification are not felt in most of the domain; GR is recovered. While, this mechanism is invoked to ensure consistency with solar system and other local tests of GR, it also has consequences on how structures grow on cosmological scales in $f(R)$. 

Modelling the growth of large scale structure in modified gravity is a more complex task than in standard gravity for the following reason. 
 In standard GR, in the absence of any particle species that may introduce anisotropic shear, the `Newtonian' gravitational potential $\Psi$, which dictates particle dynamics and appears in the Euler equation is the same as the potential $\Phi$ which appears in the Poisson equation. 
 The Laplacian in the Poisson equation is a linear operator acting on the field $\Phi$ and hence the potential can be easily solved using analytic forms or using Fourier transforms in case of periodic initial conditions. In contrast, in most MG models, the equality between $\Phi$ and $\Psi$ is violated and an extra spatial equation is necessary to evolve the two potentials simultaneously. 
If one includes the Chameleon mechanism, then one of these equations is a non-linear equation in the field and cannot be solved by simple Fourier transforms. Instead, one has to invoke more sophisticated solvers in the spatial domain. These numerical complexities do not arise in linear theory, where the main consequence of the additional degree of freedom  is that the growth rate becomes scale dependent  (e.g.\cite{Brax_2006,song_large_2007,bean_dynamics_2007,pogosian_pattern_2008}). It also does not arise in non-linear modelling, if the Chameleon mechanism is ignored.

In pioneering work, Oyaizu \cite{oyaizu_nonlinear_2008} laid down the basic numerical scheme to incorporate the effect of the Chameleon mechanism in $f(R)$ theories. The technique uses a multi-grid method for solving the equations, which involves using a hierarchy of grids with successively decreasing but pre-determined resolution. This was used \citep{oyaizu_nonlinear2_2008, schmidt_nonlinear_2009} to study the non-linear matter power spectrum  and the  distribution of haloes. Zhao et al. \cite{Zhao_2011} improved upon this by treating the high density regions with a self-adaptive mesh and this approach was used to incorporate modified gravity models into the hydrodynamic code GADGET  \citep{Puchwein_2013}. Similar numerical techniques were combined with the Zeldovich approximation and its extensions with the aim of increasing computational efficiency \citep{habib_2020}.   
There have also been efforts to model non-linear growth in $f(R)$ models using Eulerian Perturbation Theory \citep{Koyama_2009}, Lagrangian Perturbation Theory (e.g., \cite{Aviles_2017} and the COLA approach \citep{Tassev_2013,Valogiannis_2017,Winther_2017} which combines LPT evolution on large scales with N-body evolution on small scales. More recent investigations have compared numerical and approximate schemes across different types of MG models \citep{winther_modified_2015,Llinares_2018,bose2024}. 

One of the major drawbacks of N-body simulations is that they are slow. Furthermore, they employ a discrete representation of particles to represent continuous fields. The problem is exacerbated as the starting redshift increases and more particles are necessary to maintain accuracy. The use of additional non-FFT based methods to solve the extra scalar field further slows down the computation. Often, in standard gravity, semi-analytic methods such as the Press-Schechter formalism prove to be useful \cite{press_formation_1974,Sheth_2001}. These models usually employ geometric methods such as spherical or triaxial collapse as a proxy for the non-linear regime. Such geometric methods can also provide useful insights into the non-linear density-velocity divergence relation (DVDR) \citep{bilicki_velocity-density_2008,nadkarni-ghosh_non-linear_2013,nadkarni-ghosh_phase_2016}, can be useful to estimate the one-point PDF in non-standard models \citep{cataneo2021matter,mandal_one-point_2020} and can also model the evolution of halo axes ratios \citep{Lee_2005,nadkarni-ghosh_evolution_2018}. However, even implementing the spherical collapse (SC) model in modified gravity is not as simple as in standard GR. 

In this paper, we build upon the previous investigation of  Nadkarni-Ghosh \& Chowdhury (2022) \cite{Nadkarni_Ghosh_2022}; hereafter NC22, which considered spherical top-hat collapse in $f(R)$ models, but in the absence of the chameleon mechanism. One of the main aims of that paper was to understand the joint evolution of the density and velocity fields in $f(R)$ gravity and compare it to the evolution in standard GR. In standard gravity, the density and velocity divergence are proportional;  the proportionality constant is the growth rate, which is a sensitive probe of the underlying cosmology. The linear relation is a consequence of the continuity equation coupled with the linear theory solution which ignores the decaying mode. The density-velocity dynamics has been of interest since the early 1990s \citep{bernardeau_quasi-gaussian_1992,chodorowski_weakly_1997,chodorowski_recovery_1998,bernardeau_non-linearity_1999} and indeed many upcoming surveys aim to constrain the growth factor or aim to test the continuity equation \citep{zhengetal}. 

NC22 employed, a multi-step hybrid Lagrangian-Eulerian scheme to solve for the spherical collapse equations in $f(R)$ gravity. At each step, the Euler and continuity equations were solved in the Lagrangian frame by tracking shell positions.  The algebraic equations for the two gravitational potentials $\Phi$ and $\Psi$ were re-cast in terms of the variables 
$\Phi_+ =(\Phi + \Psi)/2$ and $\chi =  \Phi - \Psi$ as defined by \cite{pogosian_pattern_2008}. While $\Phi_+$ satisfies Poisson's equation, $\chi$ satisfies a non-linear equation ($\chi$ is proportional to the scalar field when mapped to quintessence models). However, if the Chameleon mechanism is ignored then the differential operator acting on $\chi$ is linear and analytic solutions are possible. In this paper, we include the Chameleon mechanism by solving the non-linear equation for $\chi$ perturbatively.  The algorithm proposed is demonstrated on the spherical collapse model, but can be extended to 3D random initial conditions as well. 

Investigations of the spherical top-hat collapse in  $f(R)$ models are not new. The seminal papers by Khoury and Weltman \citep{Khoury_2004,khoury_chameleon_2004} illustrate the chameleon mechanism using static, spherically symmetric perturbations. Brax et al. \cite{Brax_2010} elaborated on this work and generalized it to include cosmologically evolving spherical perturbations with the aim of computing the critical (linear) density of collapse, $\delta_c$, which is a key ingredient in the Press-Schechter formalism of predicting the halo mass function. 
Li and collaborators extended this to include effects of the environment, both in Eulerian and Lagrangian space, considered the effect of the nature of the random walk and compared the results with N-body simulations \citep*{li_extended_2012,Li_2012,Lam_2012,Lombriser_2013}. Most investigations with the spherical top-hat culminate in the calculation of $\delta_c$ and there have been other investigations with the same aim. \cite{borisov_spherical_2012} considered top-hat initial distributions and showed that a top-hat does not retain its top-hat shape in the presence of the chameleon screening. Instead, there is an enhancement of density near the edge of the top-hat potentially leading to shell-crossing. They considered smoothed top-hats to compute $\delta_c$, whose value in turn depended on the smoothing strength i.e. on the specific shape of the profile. In order to give a non-ambiguous estimate of $\delta_c$, \cite{kopp_spherical_2013} considered initial profiles which corresponded to average density profiles calculated using the BBKS peaks theory \citep{bardeen_statistics_1986} since they are the most relevant from an observational standpoint. \cite{Herrera_2017} also computed $\delta_c$ with different ways to approach the epoch of collapse pointing out the role of different numerical definitions of infinite density. 
\cite{lombriser_parametrisation_2016} provided a parametrized spherical collapse model as a generalization of the  linear, parametrized post-Friedmannian formalism to characterize the different effects of the modifications to gravity. Lopes et al. \cite{Lopes_2018} have investigated the spherical collapse model to compute the virial mass and turn-around radius in $f(R)$ models. In terms of techniques, the approach followed in NC22 and in this paper is similar in spirit to that of \cite{kopp_spherical_2013} and \cite{borisov_spherical_2012}. They solve the coupled continuity, Euler, Poisson and scalar field equations, the former in the Eulerian frame and the latter in the Lagrangian frame. Both use a relaxation method to solve the non-linear two-point boundary value ODE for the scalar field potential exactly. In contrast, we solve for the non-linear equation approximately, relying upon analytic solutions. Our method is approximate, but simpler to implement numerically. Furthermore, to the best of our knowledge, none of the earlier investigations of SC have explicitly presented the spatio-temporal evolution of the Compton scale, a key feature of the Chameleon mechanism. 

The paper is organized as follows. Section \ref{sec:model} defines the background $f(R)$ model used in this paper and its associated Compton wavelength. Section \ref{sec:equations} sets up the equations governing non-linear growth of perturbations. Section \ref{sec:Method} explains the method used to solve this system. Section \ref{sec:threeregimes} discusses the three regimes of evolution defined in terms of the $Q$ parameter which is the ratio of the Compton scale to the length scale of perturbation. Section \ref{sec:Results} presents the results for the spatio-temporal evolution of the Compton scale, the density and velocity fields. Section \ref{sec:Conclusion} concludes with a discussion and future outlook. 

\section{The $\lowercase{f}(R)$ model and the Compton Wavelength}
 \label{sec:model}
\subsection{The Hu-Sawicki model}


We consider a cold dark matter fluid evolving in an expanding Universe governed by an $f(R)$ model of gravity with an action of the form 
 \beq 
 S = \frac{c^3}{16 \pi G} \int d^4x \sqrt{-g} \{R + f(R)\} + S_m, 
 \label{action}
 \eeq
 where $S_m$ is a minimally coupled matter action. Here $f$ is a function of the Ricci curvature scalar $R$ defined as
\bea
 R &=& R_{\mu\nu}g^{\mu\nu} \\
 &=& \left( \partial_\alpha \Gamma^\alpha_{\mu\nu} - \partial_\mu \Gamma^\alpha_{\alpha\nu} + \Gamma^\beta_{\mu\nu} \Gamma^\alpha_{\beta\alpha} - \Gamma^\beta_{\alpha\nu} \Gamma^\alpha_{\beta\mu} \right) g^{\mu\nu},
 \eea
  where $g_{\mu\nu}$ is the metric and $\Gamma^\lambda_{\mu \nu} = \frac{1}{2} g^{\lambda \sigma} \left(\partial_\mu g_{\nu \sigma} + \partial_\nu g_{\mu \sigma} - \partial_\sigma g_{\mu \nu} \right)$. In the absence of perturbations, the spatially flat FRW background metric $g_{\alpha \beta}$ is given by 
 \beq 
 ds^2 = -c^2 dt^2 + a(t)^2 (dx^2 + dy^2 + dz^2).
 \eeq
 and the Ricci scalar for the background, denoted as $R_b$ reduces to
 \beq 
 R_b \equiv R_b(a) = \frac{12 H^2 + 6 H H'}{c^2},
 \label{Rback}
 \eeq
 where $H = \frac{{\dot a}}{a} $ is the Hubble parameter. Here $\dot{} $ and ${}'$ denote derivative w.r.t. time and $\ln a$ respectively. For a pressure-less (cold) dark matter fluid considered here, the Friedmann equation becomes,  
 \beq 
 H^2  - f_R (H H' + H^2) + \frac{c^2}{6} f + H^2 f_{RR} R'  = \frac{8 \pi  G {\bar \rho}}{3}, 
\label{backeq1}
 \eeq
 where ${\bar \rho}$ is the homogeneous  background density of dark matter, $f_R = df/dR$ and $f_{RR} = d^2f/dR^2$. 

In standard GR, the Friedmann equation is a second order equation for the scale factor $a(t)$; in $f(R)$ models, the corresponding equation is \eqnref{backeq1}, which is an equation for two unknowns $f$ and $a$. There are two approaches to solve this equation. One is to assume that the evolution of $a(t)$  or $H(t)$ is given exactly by the $\Lambda$CDM model. Then converting the derivatives of $f$ w.r.t. $R$ to derivatives w.r.t. time, gives a second order equation for $f(t)$ which is solved subject to physically realistic initial conditions. This is called the {\it designer $f(R)$} approach, in which the $\Lambda$CDM background is reproduced by construction \citep{song_large_2007,pogosian_pattern_2008}. The second approach is to assume an a functional dependence of $f$ on $R$ and choose parameters such that $\Lambda$CDM expansion history is recovered in appropriate limits. In this paper, we adopt the latter approach and focus on a particular form of $f(R)$ given by  Hu \& Sawicki \cite{hu_models_2007}: 
 \beq 
f(R) = -m^2 \frac{c_1(R/m^2)^n}{c_2(R/m^2)^n + 1}, 
\label{fofR}
\eeq
where $m^2 = \frac{H_0^2 \Omega_{m,0}}{c^2} $, with $H_0, \Omega_{m,0}$ being the Hubble parameter and the matter density parameter respectively and $c_1, c_2$ and $n$ are parameters of the model. For a flat $\Lambda$CDM expansion history, 
\bea
\label{HGR}H_{\Lambda CDM}^2(a) &=& H_0^2\left(\frac{\Omega_{m,0}}{a^3} + \Omega_{\Lambda,0} \right)\\
{\rm and} \;\; R_{\Lambda CDM}(a) &=& \frac{3H_0^2}{c^2} \left(\frac{\Omega_{m,0}}{a^3} + 4 \Omega_{\Lambda,0}\right).
\label{RGR}  
\eea

There are three constants ($c_1, c_2$ and $n$) in \eqnref{fofR} that need to be fixed. Following \citep{hu_models_2007}, we impose the additional physical condition that at early epochs, when $R \rightarrow \infty$, the dynamics should reduce to that given by a $\Lambda$CDM model. This gives the condition 
\beq 
\lim_{R \rightarrow \infty} f(R) = f_{\Lambda CDM}.  
\eeq
For a $\Lambda$CDM model, the action is written as $ S = \frac{c^3}{16 \pi G} \int d^4x \sqrt{-g} \{R - 2 \Lambda\} + S_m$, giving the identification 
\beq 
f_{\Lambda CDM}= -2 \Lambda = -\frac{16 \pi G \rho_\Lambda}{c^2} = - \frac{6 H_0^2 \Omega_{\Lambda,0}}{c^2},
\label{flcdm}
 \eeq
where, $\Lambda$ is the cosmological constant and $\rho_\Lambda = \Lambda c^2/(8 \pi G) = \rho_{\Lambda, 0}$ is the constant energy density of the vacuum component.  
Evaluating expression \eqnref{fofR} for large $R$, gives 
\beq 
\lim_{R \rightarrow \infty} f(R)  \approx - m^2 \frac{c_1}{c_2} = \frac{H_0^2 \Omega_{m,0}}{c^2} \frac{c_1}{c_2}. 
\label{flargeR}
\eeq
Equating \eqnrefs{flcdm} and \eqnref{flargeR} gives the relation 
 \beq 
\frac{c_1}{c_2} \approx 6 \frac{\Omega_{\Lambda,0}}{\Omega_{m,0}}.
\label{c1c2rel}
\eeq  
Denote the value of $f_R$ at the present time as $f_R(a_0) = f_{R0}$, then from  \eqnref{fofR}, we have 
\beq 
f_{R0} \approx -n \frac{c_1}{c_2^2} \left(\frac{12}{\Omega_{m,0}} -9\right)^{-n-1}, 
\eeq
which relates the $c_1, c_2$ and $n$ to $f_{R0}$. Imposing \eqnref{c1c2rel} gives two remaining free parameters: $n$ and $f_{R0}$ for the $f(R)$ model. 
In this work, we choose the parameters: $\Omega_{m,0} = 0.32, \Omega_{\Lambda,0} =0.68 , f_{R,0} = -10^{-6}, n=1$ (same as NC22). Certain features of the background evolution are worth noting. In the designer $f(R)$ model, the background $\Lambda$CDM model is recovered at all times by construction. Although,  the parameters of the $f(R)$ model are chosen such that the action reduces to its $\Lambda$CDM value at early times, it does not ensure that $w= -1$ at all times. It is therefore fair to ask the question: given a $f(R)$ model, how closely does it follow the $\Lambda$CDM model at late times ? After all, at the background level, phenomenologically, the $\Lambda$CDM model is a good description of the observed acceleration of the Universe.  This can be analysed by substituting \eqnref{fofR} in \eqnref{backeq1} and solving the resultant along with \eqnref{Rback} for the variables $H$ and $R$. Following \cite{hu_models_2007}, we can re-cast these equations in terms of first order ODEs for the variables\footnote{In \cite{hu_models_2007} and also in NC22, the action was defined with $c=1$ and the corresponding $R$ and $f(R)$ were defined to have dimensions of $T^{-2}$. In this paper, the factor of $c$ is included in the definition of the action and $R$ and $f(R)$ have dimensions of $ L^{-2}$. }
\bea
y_H &=& \frac{H^2 }{H_0^2 \Omega_{m,0}} -a^{-3}  \\
y_R &=& \frac{R c^2}{H_0^2 \Omega_{m,0}} -3 a^{-3}. 
\eea
These variables have the advantage that their values are constant for the $\Lambda$CDM model given by $y_{H,\Lambda CDM} = \Omega_{\Lambda,0}/\Omega_{m,0}$ and $y_{R,\Lambda CDM} = 12 \Omega_{\Lambda,0}/\Omega_{m,0}$.
In terms of these variables, \eqnrefs{Rback} and \eqnrefbare{backeq1}  become 
\bea
 \label{eqnyh} y_H'&=& \frac{1}{3} y_R - 4 y_H, \\
\label{eqnyr} y_R'&=& \frac{9}{a^3} -\frac{1}{{\tilde f}_{{\tilde R}{\tilde R}}(y_H + a^{-3})} \left[  y_H  - {\tilde f}_{\tilde R} \left(\frac{y_R}{6} -y_H -\frac{1}{2 a^3}\right) +\frac{\tilde f}{6}\right], 
\eea
where 
\beq 
{\tilde f} = \frac{f}{m^2} \;\;\; {\rm and} \;\;\; {\tilde R} = \frac{R}{m^2}.
\eeq
The choice of the initial epoch is arbitrary and GR is assumed to be valid at that time. The effective equation of state\footnote{$w_{eff}$ is defined through the relation $H^2 = H_0^2\left(\frac{\Omega_{m,0}}{a^3} + \Omega_{de,0} \exp\left[{-\int_{a}^1 \left\{1+ w_{eff}(u)\right\} \frac{du}{u}}\right]\right)$, $u$ being a dummy variable. } in terms of these variables is 
\beq 
1+ w_{eff}(z) = -\frac{1}{3} \frac{y_H'}{y_H}.
\label{eqnofstate}
\eeq 
The question is: how does $w_{eff}(z)$ evolve with redshift, in particular, does it reproduce $w_{{\Lambda}{\rm CDM}} = -1$ at late times ? 
The earlier work, NC22 investigated this question. It turns out that how closely it follows $-1$ depends upon the starting redshift for the evolution of \eqnrefs{eqnyh} and \eqnrefbare{eqnyr}. The earlier the starting epoch, the more closely it follows $-1$ at late times (see figure 1 of NC22). Thus, one would, in principle, like to start evolution as early as possible. However, the solutions to $y_H$ and $y_R$ and consequently $w_{eff}$ are oscillatory functions, whose frequency of oscillations increases with increasing redshift. Thus, starting the evolution very early is computationally intensive because of numerical stiffness.
In NC22, we performed an eigenvalue analysis of the $\{y_H,y_R\}$ dynamical system so as to understand the oscillatory behavior  (it is realated to the magnitude of complex eignevalues). This analysis allows one to predict an epoch where the amplitude of oscillations starts to decrease (denoted as $a_{trans}$, see figure 2 of NC22). In order to balance the stiffness issue with the effects departure from $\Lambda$CDM, we follow the approach of evolving the perturbations in standard gravity until a certain epoch $a_{switch} \leq a_{trans}$ after which the $f(R)$ evolution takes over. In NC22, $a_{switch} = 0.1$ and we follow the same in this paper.

\subsection{The Compton  wavelength}

In this paper, the action given by \eqnref{action} and the subsequent evolution equations are represented in the Jordan frame. It is possible, via a conformal transformation of the coordinates, to recast this theory to have a standard Einstein-Hilbert action with a scalar field which introduces a non-minimal coupling in the matter sector. This is the Einstein frame 
\citep{magnano_physical_1994,chiba_1r_2003,sotiriou_fr_2010}. Although, the original formulation of the Chameleon mechanism was developed in the Einstein frame \citep{Khoury_2004,khoury_chameleon_2004}, the mapping between the two frames identifies the scalar field with $f_R$. The associated Compton scale gets defined as \citep{hu_models_2007,Navarro_2007}
\beq
\lambda_C = 2 \pi \left[ \frac{1}{3} \left(\frac{1 + f_R}{f_{RR}} -R \right) \right]^{-1/2} \approx 2 \pi \sqrt{3 f_{RR}}.
\label{Comptondef}
\eeq
The second approximate equality follows from the fact that $f_R<<1$ and ${f_{RR}}^{-1} >>R$ throughout the evolution. 
The comoving Compton wavelength $x_C$ is defined as 
\beq 
x_C = \frac{\lambda_C}{a}, 
\eeq 
but the equations depend on the `reduced' comoving Compton wavelength ${\bar x}_C$ defined as 
\beq 
{\bar x}_C = \frac{x_C}{ 2\pi} \approx   \frac{ \sqrt{3   f_{RR}}}{a}.
\label{Comptondef2}
\eeq 
In the rest of this paper, we refer to ${\bar x}_C$ as the Compton scale. In the Einstein frame, as the perturbation evolves, the scalar field which is coupled to the matter density evolves and the value of the potential of the field changes, which in turn changes its mass and Compton wavelength. In the Jordan frame, the change in density causes a change in the Ricci curvature scalar $R$ which changes the value of the Compoton wavelength \citep{hu_models_2007}.
We define the background value of the Compton scale as 
\beq 
{\bar x}_{C,b} \approx  \left.\frac{ \sqrt{3   f_{RR}}}{a} \right|_{R = R_b}, 
\eeq
where $R_b$ is given by \eqnref{Rback}. In principle, $H(a)$ in \eqnref{Rback} should be evaluated by solving \eqnref{backeq1} with the Hu-Sawicki form for $f(R)$. However, 
for the value of $f_{R0}=-10^{-6}$ considered in this work, $H(a)$ evaluated this way is expected to agree with its $\Lambda$CDM value given by \eqnref{HGR} to within 0.1\% (from  fig.1 of NC22, the deviation of $w_{eff}$ was less than 0.1\%). Thus, practically, we evaluate the above expression with $R_b = R_{{\Lambda}CDM}$ given by \eqnref{RGR}.

\section{Equations governing perturbations }
 \label{sec:equations}
 We consider only scalar perturbations of the metric evolving in the background cosmology described above. The perturbations are only  in the matter sector with the density contrast given by $\delta = \rho/{\bar \rho}-1$, where $\rho$ is the density of the perturbations and ${\bar \rho}$ is the background density. ${\bf x} = \{x_1, x_2, x_3\}$ denotes the comoving coordinate and ${\bf v} = a {\dot {\bf x}}$ is the peculiar velocity. In the conformal Newtonian gauge the metric has the form 
  \beq 
 ds^2 = -c^2\left(1 + \frac{2 \Psi}{c^2}\right)dt^2 + a^2\left(1 - \frac{2\Phi}{c^2}\right) \left(dx_1^2 + dx_2^2 + dx_3^2 \right).
 \label{perturbedmetric}
 \eeq 
 Here, $\Psi$ corresponds to the Newtonian potential, which dictates particle motions  and $\Phi$ is the `curvature' fluctuation. 
  In standard gravity with $\Lambda$CDM, on sub-horizon scales, equations of Newtonian hydrodynamics (continuity, pressure-less Euler) coupled with Poisson's equation describe the evolution of perturbations fairly well. These equations are effectively obtained from the sub-horizon limit of Einstein's equations ($ck <<aH$) and the equation governing the conservation of the stress-energy tensor. In this limit, when the action is the standard GR action, the off-diagonal space-space part of Einstein's equations gives  
\beq
\nabla^2 (\Phi  - \Psi) = 12 \pi G a^2  (\bar \rho + \frac{\bar p}{c^2}) \sigma, 
\eeq 
where $\bar \rho$ and $\bar p$ are the background density and pressure, $\sigma$ is the dimensionless, anisotropic (shear) stress term of the stress-energy tensor that describes the fluid \cite{ma_cosmological_1995}. When the cosmological fluid consists only of dark matter and non-clustering  dark energy, no shear  
is generated and the two potentials are equal, i.e. $\Phi = \Psi$\footnote{If the fluid has a free-streaming relativistic component such as photons or neutrinos, then $\sigma \neq 0$ because as the particles move out from high density regions into low density regions they generate a directionality in the velocity field which gives rise to the shear stress.  This gives $\Phi \neq \Psi$ even for standard Einsteinian gravity. Accurate tracking of perturbations in such situations requires general relativistic codes such as those developed by \citep{Fidler_2015,Adamek_2016,Adamek_2016b}.}.
However, in $f(R)$ gravity, since the action $S$ changes, the Einstein equations which are obtained by varying the action (in the standard metric formalism) also change \citep{sotiriou_fr_2010}. Now, the difference $\Phi-\Psi$ is related to the new term in the action i.e., $f(R)$. In particular, it is proportional to the perturbation in $f_R$ ($\delta f_R$) as \footnote{$\delta f_R$ and $\delta R$ denote perturbations in the scalar field $f(R)$ and the Ricci scalar respectively. In the chameleon mechanism, they are dependent on the density contrast $\delta$, but the notation should not be confused: $\delta f_R \neq \delta \cdot f_R$ and $\delta R \neq \delta \cdot R$} 
 \beq 
 \Phi-\Psi  =  c^2 \delta f_R. 
 \label{deltafR_1}
 \eeq
 
 In modified gravity models, in addition to the  sub-horizon approximation, we separately impose the quasi-static approximation\footnote{In standard $\Lambda$CDM gravity, it can be shown that imposing the sub-horizon approximation, automatically imposes the quasi-static approximation. This is not true in other non-standard dark energy models and the two approximations have to be invoked separately.}, wherein the potentials $\Phi$ and $\Psi$ vary on longer time scales as compared to the expansion time scale $H^{-1}$. In this limit, the Einstein equations for the potentials are \citep{oyaizu_nonlinear_2008,schmidt_nonlinear_2009,borisov_spherical_2012},
 \bea
 \label{deltafReq}\nabla_x^2 \delta f_R &=& \frac{a^2 \delta R}{3} -  \frac{H^2 a^2 \Omega_m \delta}{c^2},\\
 \nabla_x^2 \Psi &=&  2 H^2 a^2 \Omega_m \delta - \frac{a^2 \delta R c^2}{6}, 
 \label{psieq}
 \eea
 where $\delta R$ is the perturbation to the Ricci scalar $R$. In a spatially varying density perturbation, $R$ will be a function of ${\bf x}$. From the definition of $f_{RR}$, we have 
  \bea
 \delta f_R &=& f_{RR}\delta R  \label{deltaR_eq1} \\
 {\rm or}\;\; \delta R &=& f_{RR}^{-1} \delta f_R  =   \frac{\Phi-\Psi}{f_{RR} c^2}. 
 \eea

We note that this is an exact expression for $\delta R$. In NC22, where the Chameleon mechanism was ignored, we assumed that $f_{RR}$ is evaluated only at the background $R_b$ and is a purely temporal function i.e.  $f_{RR} \equiv \left.f_{RR}\right|_{R_b} \equiv  f_{RR}(a)$. However, when the Chameleon screening is included, $R$ becomes a spatio-temporal function and $f_{RR}$ varies spatially, in addition to its background temporal evolution. This is true for all scales, in general. We would also expect the change to be small on large scales where the density contrast $\delta$ is small and more on small scales where $\delta$ is larger. 
However, as we shall see, the dynamics is dictated not only by the strength of the perturbation, but also by the ratio of the comoving Compton scale to the scale of the perturbation.

 Following \cite{pogosian_pattern_2008} we define the variables 
  \bea
 \label{chidef} \chi &=&  \Phi - \Psi \\
 \label{Phidef} \Phi_+ & = & \frac{\Phi + \Psi}{2}. 
 \eea 
Substituting \eqnref{deltafR_1}, \eqnref{deltaR_eq1}, \eqnref{chidef} and \eqnref{Phidef} in \eqnref{deltafReq} and  \eqnref{psieq} 
and adding  the appropriate limit of the conservation of the stress-energy tensor gives the system 
 \bea 
 \label{conteq}\frac{\partial \delta}{\partial t} + \left(\frac{{\bf v}}{a} \cdot \nabla_x\right) \delta &=& -\frac{(1+\delta)}{a} (\nabla_x \cdot {\bf v})\\
 \label{euler}\frac{\partial {\bf v}}{\partial t} + \left(\frac{{\bf v}}{a} \cdot \nabla_x\right) {\bf v} +  H {\bf v}& =& -\frac{1}{a} \nabla_x \Psi\\
 \label{phiplus}\nabla_x^2 \Phi_+& =&\frac{3}{2}  H^2  a^2 \Omega_m \delta \\
 \label{chi}\nabla_x^2 \chi  - \frac{ a^2}{3  f_{RR}({\bf x},a)}  \chi &=&- H^2  a^2 \Omega_m \delta.
 \eea
Substituting from \eqnref{Comptondef2}, we can re-write the last equation as 
\noindent 

\beq
\nabla_x^2 \chi  - \frac{1}{{\bar x}_C^2({\bf x},a)}  \chi =- H^2  a^2 \Omega_m \delta.
\label{chi2}
\eeq
These equations are to be solved for the coupled variables $\delta$, ${\bf v}$, $\Phi_+$ and $\chi$ given initial spatial profiles for the $\delta$ and ${\bf v}$.  For 3D Gaussian initial conditions, the spatial solutions are subject to periodic boundary conditions. For spherical symmetry the boundary conditions are given in Section \ref{sec:Method}.

In this system, the dependence of $f_{RR}$ or ${\bar x}_C$ on the position is indicated explicitly. It is possible to approximate \eqnref{deltaR_eq1} as 
\beq 
\delta f_R \approx \left.f_{RR}\right|_{R = R_b} \delta R + \mathcal{O}(\delta R^2) 
\eeq
and consequently drop the position dependence of the Compton scale in \eqnref{chi2}. This is equivalent to ignoring the Chameleon mechanism as was done in NC22. In general, as the density perturbation grows, $R$ changes as a function of both time and space and consequently so does $f_{RR}$ and the related Compton scale. In this work, we track the change in $R$ and evaluate $f_{RR}$ at the appropriate $R$. Given the form of the metric in \eqnref{perturbedmetric}, the expression for $R$ is

\begin{align}
  R =  \frac{6 \left[H'H+2 H^2\right]}{V_\Psi} &+ 
 \frac{c^2 \left[V_\Psi \left(2 V_\Psi \partial_{x}^2 (\Phi) -V_\Phi \partial_{x}^2(\Psi) \right)+V_\Phi \left(\partial_{x}(\Psi)\right)^2\right]}{a^2 V_\Phi^2 V_\Psi ^2} \nonumber \\ 
&+ \frac{c^2 \left[ V_\Phi  \partial_{x}(\Psi) \partial_{x}(\Phi) + 3 V_\Psi^2 \left(\partial_{x}(\Phi) \right)^2\right]}{a^2 V_\Phi^3V_\Psi},
\label{eqnforR}
\end{align}
where $V_\Psi = c^2+2\Psi$ and $V_\Phi = c^2 - 2\Phi$.  Note that in the absence of perturbations, it reduces to \eqnref{Rback}. \\


\section{Method of solution}
\label{sec:Method}
 \subsection{Algorithm}
 The hybrid Eulerian-Lagrangian scheme used to solve this system of equations is similar to that discussed in NC22, except for a modification for the solution of $\chi$. We briefly review the basic method here. The set of equations given is evolved using standard GR equations until $a=a_{trans}$ after which the system is evolved in $f(R)$ gravity. The interval of interest from $a_{trans}$ to $a_{final}$ is divided into $N_t$ intervals. The density is assumed to stay constant in between the time-step. At each time step, first the potentials $\Phi$ and $\Psi$ are computed given the density at the start of the step. The system is then evolved in time for the duration of the step using these values of potentials. The updated density and velocity are computed at the end of the time step. The algorithm is as follows. Let $a_n$ and $a_{n+1}$ denote the starting and stopping epochs of the $n$-th interval. The density, $\delta({\bf x}, a_n)$,  and peculiar velocity, ${\bf v}({\bf x},a_n)$, are known at the start of the interval. The evolution between each time step consists of two operations as follows:
\begin{itemize}
 \item  Step I - {\it Solve the spatial equations for the potentials in the Eulerian frame. }\\
  The spatial equations \eqnref{phiplus} and \eqnref{chi} are solved for the fields $\Phi_+$ and $\chi$ assuming that the density stays constant over the time interval. 
 \bea
 \label{phiplus_an}\nabla_x^2 \Phi_+& =&\frac{3}{2}  H^2  a^2 \Omega_m \delta({\bf x}, a_n), \\
  \label{chi_an}\nabla_x^2 \chi  - \frac{ a^2}{3  f_{RR}({\bf x},a_n)}  \chi &=&- H^2  a^2 \Omega_m \delta({\bf x}, a_n).
 \eea
In this paper we will consider smooth spherical top-hat perturbations. Imposing spherical symmetry, the equations simplify to 
 \bea
\label{phiplus_sph}\frac{\partial^2 {\Phi_+}}{\partial x^2} + \frac{2}{x} \frac{\partial{\Phi}_+}{\partial x} &=& \frac{3}{2} H^2 a^2 \Omega_m(a) \delta(x,a_n), \\
\label{chieq_sph}\frac{\partial^2 {\chi}}{\partial x^2} + \frac{2}{x} \frac{\partial{\chi}}{\partial x} - \frac{ a^2}{3  f_{RR}({\bf x},a_n)}  \chi  &=& - H^2 a^2 \Omega_m(a) \delta(x,a_n),  
\eea 
to be solved subject to boundary conditions 
 \bea
\label{bc1} \Phi_+(x\rightarrow \infty) = 0,  \;\;\; \left.\frac{\partial \Phi_+} {\partial x}\right|_{x=0} &=& 0, \\
\label{bc2}  \chi(x\rightarrow \infty) =0,  \;\;\;  \left.\frac{\partial \chi}{\partial x}\right|_{x=0} &=&0.
 \eea
 If the Chameleon mechanism is ignored, then $f_{RR}$ is not a function of space and the solutions for $\Phi_+$ and $\chi$ have analytic solutions (see Appendix C of NC22)  When the Chameleon mechanism is accounted for, the solution for $\chi$ is given by the perturbative scheme outlined in section \ref{Chisoln}. 
\item Step II - {\it Solve the Euler and continuity equations for density and velocity in the Lagrangian frame.} \\
 In the case of spherical symmetry, the radially varying density perturbation is modelled as a series of spherical shells, each with a different density and peculiar velocity.  Let the origin be the centre of the sphere. The Lagrangian coordinate of any shell is taken to be its initial comoving coordinate  ${\bf q}$ at the start of each step. The Eulerian (physical) coordinate at any later epoch is given by 
\beq
 x\equiv x (q,a) = A(q, a) q, 
 \label{xdef}
\eeq
where $A(q, a)$ is the scale factor of the shell at $q$. The peculiar velocity is given by 
 \beq 
 v(q,a) = a {\dot A} q =  a A'(q, a) H(a) q.
 \label{pecvelLag}
 \eeq
  With these definitions, the Euler equation \eqnref{euler} in Lagrangian variables becomes
 \beq 
\label{eqnforA} A'' + \left(2 - \frac{3}{2} \Omega_m(a) \right) A' = -\frac{1}{q} \nabla_q {\tilde \Psi}(q,a), \\
\eeq
where ${\tilde \Psi} = \Psi/H^2$ is obtained from Step I. By definition, $A=1$ for all shells at the start of every step. The initial peculiar velocity at the start of the step $v(q, a_n)$ sets the initial conditions for $A'$ at the start of each step through \eqnref{pecvelLag}.

The density at the end of the step at $\delta(q, a_{n+1})$ is related to the density at the start of the step $\delta(q, a_{n}) $ as
\beq 
1+\delta(q, a_{n+1} ) = \left [1+ \delta(q,a_{n})\right] \frac{q^2 dq}{x^2dx} =  \frac{\left[1+ \delta(q, a_n)\right]}{A^3 \left|1+ \frac{q}{A} \frac{dA}{dq}\right|}. 
\label{Lagdensity} 
\eeq
 The spherically averaged density is defined as 
\beq
\Delta(x) = \frac{3}{x^3} \int_0^x  \delta(x,a) x^2 dx. 
\label{Deltadef}
\eeq
It satisfies the condition 
\beq 
1  + \Delta(q,a_{n+1} ) = \frac{(1 +\Delta(q, a_n)) A^3_{init}}{A^3}.
\label{Deltaconserve}
\eeq
Having solved for $A(q, a_{n+1})$, the density and velocity for the next step are computed through \eqnref{pecvelLag} and \eqnref{Lagdensity}. These are known as functions of the Lagrangian coordinate at $a_n$ and need to be interpolated and re-defined on an equi-spaced grid in the final Eulerian space. At the start of each time step, one needs to define a new Lagrangian coordinate, thus $q_{n+1} \neq q_{n}$. We have denoted it generally as $q$ in the expressions above. 
\end{itemize}
 
 \subsection{Perturbative solution for the $\chi$ field} 
 \label{Chisoln}
This section  describes the perturbative solution to \eqnref{chi_an}.
 In the presence of perturbations, the Ricci scalar becomes $R = R_b + \delta R$, where $R_b \equiv R_b(a)$ is the Ricci scalar of the background and $\delta R$ is the modification due to the perturbation. In principle, $\delta R$ and hence $\left.f_{RR}\right|_{R + \delta R}$, depends on $\chi$ which makes \eqnref{chi_an} an implicit non-linear equation. However, we will assume that the dominant contribution to $\delta R$ is from the estimate of $\chi$ in the absence of Chameleon. The corrections to $\chi$ are systematically calculated using this value of $\delta R$. We assume the ansatz 
 \beq 
\chi = \sum _{n=1}^\infty \epsilon^n \chi^{(n)},  
\label{chiexpansion}
\eeq
where $\epsilon$ is a book-keeping parameter, whose magnitude is of the order of $\delta({\bf x}, a_{n})$ or $\delta R$. Expanding $f_{RR}$ around $R = R_b$ gives,  
  \bea 
 f_{RR}(R)   &=& \left.f_{RR}\right|_{R_b} + \sum_{n=1}^{\infty} \left.\frac{d f^{n+2}}{d R^{n+2}} \right|_{R_b} \frac{ (\epsilon  \delta R)^n}{n!}\\
 &=&  \left.f_{RR}\right|_{R_b} \left(1 + \sum_{n=1}^{\infty}  F_n  (\delta R)^n \epsilon^n \right), 
 \label{Taylor}
 \eea
 where $F_n$ is defined as 
 \beq
 F_n = \frac{1}{n! \left.f_{RR}\right|_{R_b} } \left.\frac{d f^{n+2}}{d R^{n+2}} \right|_{R_b}.
 \eeq
 Substitute \eqnref{chiexpansion} and \eqnref{Taylor} in \eqnref{chi_an} and identify $\frac{ a^2}{3  \left.f_{RR}\right|_{R_b}}  = \frac{1}{{\bar x}_{C,b}^2}$, the Compton wavelength for the background. This gives
\beq
 \nabla_x^2 \left( \sum _{n=1}^\infty \epsilon^n \chi^{(n)}\right) -\frac{1}{{\bar x}_{C,b}^2} \left( \sum _{n=1}^\infty \epsilon^n \chi^{(n)}\right) \cdot  \left(1 + \sum_{n=1}^{\infty}  F_n  (\delta R)^n \epsilon^n \right)^{-1} = 
 - H^2  a^2 \Omega_m \delta({\bf x}, a_{n}).
\eeq
Equating order-wise, gives at first order, 
 \beq
 \nabla_x^2 \chi^{(1)}  - \frac{1}{{\bar x}^2_{C,b}} \chi^{(1)} = - H^2  a^2 \Omega_m \delta({\bf x}, a_n).
  \label{firstO}
 \eeq
 After solving for $\chi^{(1)} $ and $\Phi_+$, $\delta R$ is computed as 
 \beq 
 \delta R = R - R_b(a_n), 
 \eeq
 where $R$ is given by \eqnref{eqnforR} and $R_b(a_n)$ is the background value of $R$ at $a=a_n$.  The $\Phi$ and $\Psi$ appearing in \eqnref{eqnforR} are evaluated from $\Phi_+$ and $\chi = \chi^{(1)}$. 
Equating the higher order $\epsilon$ terms, gives equations for the correction terms as:
\begin{align}
   \nabla_x^2 \chi^{(2)}  - \frac{1}{{\bar x}_{C,b}^2} \chi^{(2)} &= \frac{1}{{\bar x}_{C,b}^2} F_1 (\delta R) \chi^{(1)} \\
 \nabla_x^2 \chi^{(3)}  -\frac{1}{{\bar x}_{C,b}^2}  \chi^{(3)} &= \frac{1}{{\bar x}_{C,b}^2}  \left\{ \left(F_1^2 - F_2\right)  (\delta R)^2  \chi^{(1)} - F_1 (\delta R)  \chi^{(2)} \right\} \\
\nabla_x^2 \chi^{(4)}  -\frac{1}{{\bar x}_{C,b}^2}  \chi^{(4)} &= -  \frac{1}{{\bar x}_{C,b}^2} 
 \left\{    \left(F_1^3 - 2F_1F_2 + F_3\right) (\delta R)^3 \chi^{(1)}  +\left(F_2 - F_1^2\right) (\delta R)^2  \chi^{(2)}+ 
 F_1 (\delta R)\chi^{(3)} \right\}
 \end{align}
  The structure of the equations is such that the  higher order source terms are combinations of lower order solutions and their form can be easily generated using a symbolic software like Mathematica \citep{math}.  In this paper, we stop at a fourth order expansion in $\epsilon$. The boundary conditions given by \eqnref{bc1} and \eqnref{bc2} translate at each order to
  \beq
\chi^{(n)}(x\rightarrow \infty) =0, 
 \;\;\;  \left.\frac{\partial \chi^{(n)}}{\partial x}\right|_{x=0} =0 \;\;\; {\rm for} \; n =1,  2, 3, 4. 
 \eeq
For the spherically symmetric perturbations considered in this paper, the $\nabla^2$ operator simplifies to that in \eqnref{chieq_sph} and the solutions can be constructed using the analytic forms of NC22. Appendix \ref{app:1} presents the convergence test of this approach implemented on such profiles and appendix \ref{app:2} presents the convergence test of the algorithm for the temporal evolution described in the previous sub-section. 

It is worth noting that the above framework for the iterative solution is general and can be easily extended to 3D perturbations with random initial conditions in a periodic box. Each term $\chi^{(n)}$ can be solved using FFTs. The temporal equations for $\delta$ and ${\bf v}$ can be solved using perturbative techniques in the Eulerian frame \citep{bernardeau_2002} or in the Lagrangian frame (for e.g., \citep{nadkarni-ghosh_extending_2011,nadkarni-ghosh_modelling_2013}).

 
\section{Three regimes of evolution}
\label{sec:threeregimes}

The variable $\chi$ is related to the extra scalar degree of freedom and the equation for its evolution i.e., \eqnref{chi2} consists of two terms $\nabla_x^2 \chi$ and $\frac{1}{{\bar x}_C^2} \chi$ on the l.h.s and the source term proportional to $\delta$ in the r.h.s. There are two aspects to understanding the evolution of $\chi$. One is the relative importance of the two terms on the l.h.s. and the second is how this relative importance may change when the Compton scale ${\bar x}_C$ changes due to the evolution of the perturbation (Chameleon screening). 
In this paper, we consider smooth, compensated top-hat perturbations whose profile is characterized by a co-moving scale $x_{top}$ (details in Appendix \ref{app:0}). $x_{top}$ is a fixed quantity\footnote{As a spherical overdensity evolves, it will compactify but the comoving scale will not change by orders of magnitude for the quasi-linear evolution considered here} and this scale dictates the magnitude of the $\nabla_x^2 \chi$ term. 
Thus, the relative importance of the two terms on the l.h.s. can be quantified by a parameter $Q$ defined as 
\beq
Q = \frac{{\bar x}_C}{x_{top}}. 
\label{Qdef}
\eeq
In the absence of the Chameleon mechanism, ${\bar x}_C$ is a purely time-dependent quantity and so is $Q$. In the presence of the Chameleon screening, the Compton scale becomes a function of both time and space. More explicitly, we can write ${\bar x}_C = {\bar x}_{C,b}(a) + \Delta {\bar x}_{C}({\bf x}, a)$, where $ \Delta {\bar x}_{C}({\bf x}, a)$ is the change in the Compton scale due to the perturbation. $\Delta {\bar x}_{C}({\bf x}, a)$ depends on $\delta R$, which depends on $\chi$, which in turn depends on $\delta$. Thus, the parameter $Q$ is implicitly determined by the matter density $\delta$. Similarly, the solution for $\chi$ has an explicit dependence on $\delta$ through the source term, but an implicit dependence through the term $\frac{1}{{\bar x}_C^2} \chi$. Indeed, tracking this dependence on $\delta$ is the essence of the Chameleon screening and the perturbative solution described in the previous section does precisely that. Based on $Q$, we define three regimes of the dynamics:

\begin{enumerate}
\item  The strong-field regime ($Q>>1$): In this regime, the range of the fifth-force (${\bar x}_C$) is much larger than the length scale of the perturbation and \eqnref{chi2} for $ \chi$ reduces to a Poisson-like equation for $\chi$:  $\nabla^2 \chi \propto \delta$. This is similar to the equation for $\Phi_+$ and the effect of the modification is that the force is 4/3 times its value in standard GR. Often, this effect is encapsulated by defining a modified Newton's constant $G_{eff} = \frac{4}{3} G$ (e.g., \cite{pogosian_pattern_2008}). In this regime, the system is indifferent to the presence or absence of the Chameleon mechanism. 
\item The weak-field regime ($Q <<1$): Here, the force range is much smaller than the length scale of the perturbation. The naive expectation is that the fifth-force is unimportant and the dynamics will reduce to GR. While $\chi$ is small, it is non-zero and the solution for $\chi$ can be approximated as $H^2 a^2 \Omega_m {\bar x}_C^2 \delta$. Consequently, the extra force $\nabla_x \chi$ is proportional to $\nabla \delta$. We expect to see effects of the modification only in regions where the density changes over a length scale much smaller than the Compton wavelength. In this regime, it is possible to obtain density enhancement at the top-hat edge even in the absence of the Chameleon screening as was demonstrated in NC22. 

\item The intermediate regime ($Q \sim 1$):  In this regime, both terms  in \eqnref{chi2} are important and the effect of the Chameleon screening can be seen in the dynamics of the perturbations. 
\end{enumerate}
Note that in the presence of Chameleon screening, $Q$ is a time as well as space dependent quantity and the inner and outer parts of a perturbation can potentially be in different regimes. 


\section{Numerical runs and results}
\label{sec:Results}

\begin{figure}
\centering
\includegraphics[width=16cm]{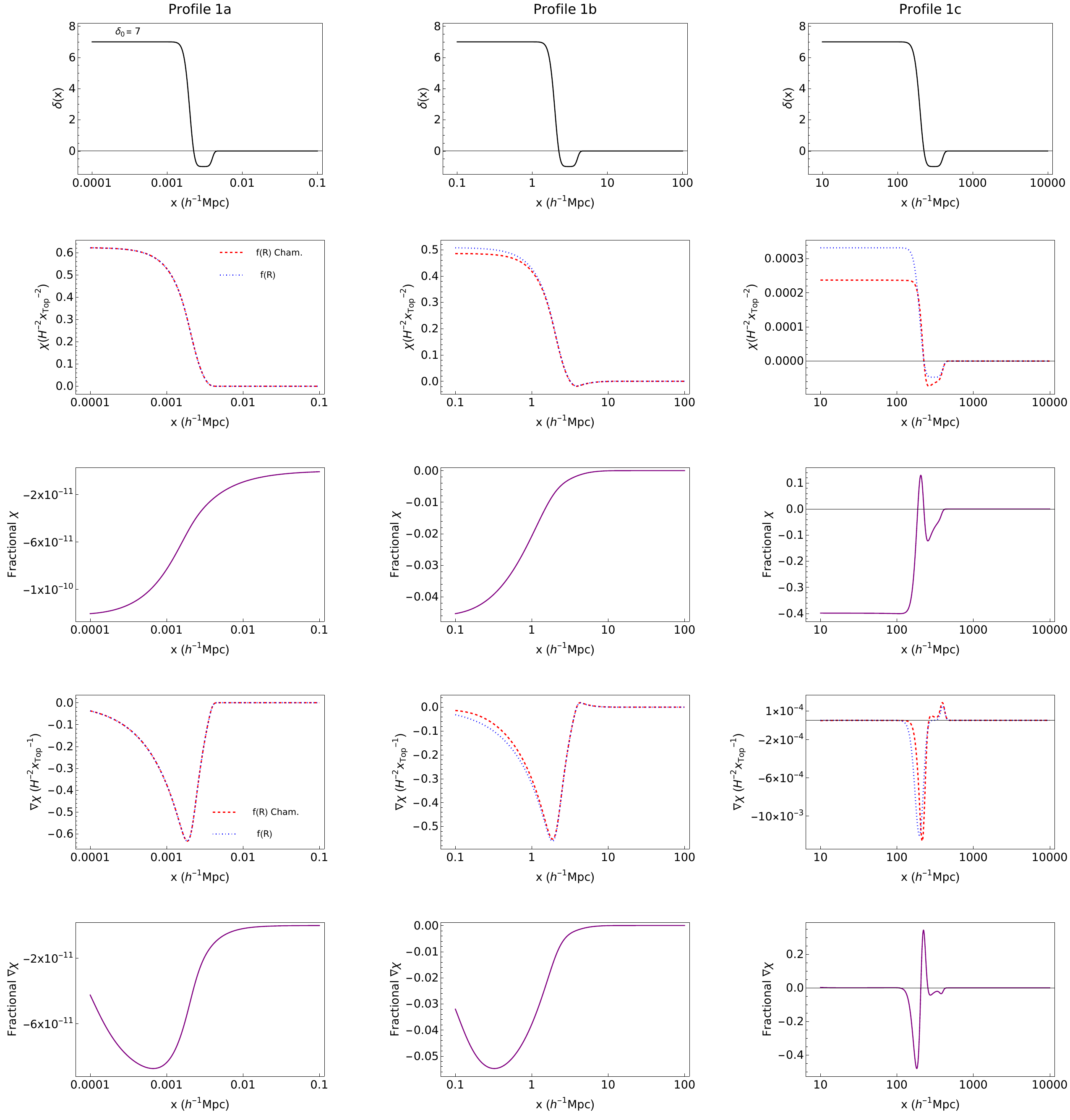}
\caption{The spatial solutions for the field $\chi$ scaled by $(Hx_{top})^{-2}$ and its gradient $\nabla \chi$, also scaled by the same factor, with (red, dashed) and without (blue, dotted) the Chameleon mechanism. Fractional $\chi$ is the difference  between $\chi$ and $\chi^{(1)}$ divided by the maximum value of $|\chi|$ over the radial grid. Here, $\chi$ and $\chi^{(1)}$ are the solutions with and without the screening respectively. Similarly, fractional $\nabla \chi$ is the difference between $\nabla \chi$ and $\nabla \chi^{(1)}$ divided by the absolute maximum value of $|\nabla \chi|$. Profile 1a is in the strong field limit and hence unaffected by the inclusion of the Chameleon mechanism, profile 1c is in the weak field limit and is affected only at the edge of the top-hat where the density gradient is large. Profile 1b shows differences in the solution with and without screening throughout the profile.  }
\label{Staticsoln}
\end{figure}

\begin{figure}
 \includegraphics[width=16cm]{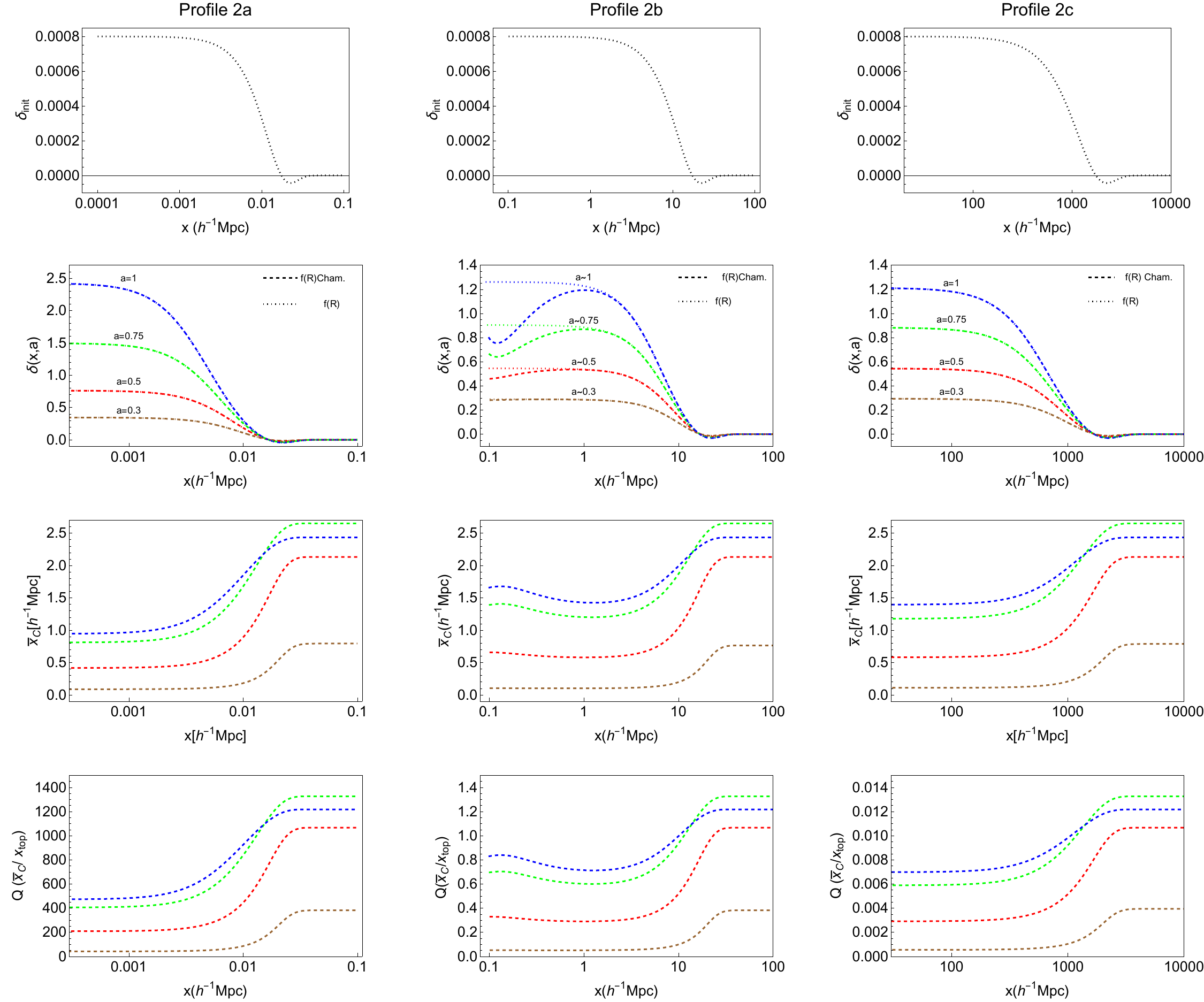}
 \caption{The Chameleon mechanism: spatio-temporal evolution of the density field and associated Compton scale. The initial profile (first panel), the evolved density (second panel), the scaled Compton wavelength (third panel) and the $Q$ factor (fourth panel) plotted for three different initial profiles. Profile 2a stays in the strong field regime. The Chameleon mechanism does not manifest in this regime. Profile 2c stays in the weak field limit and again does not exhibit the Chameleon screening. Only profile 2b shows a change in behaviour due to the Chameleon screening. The shape of the top-hat is changed and a monotonically decreasing profile is no longer monotonic.  }
 \label{Compton}
 \end{figure}
 
\begin{figure}
\includegraphics[width=16cm]{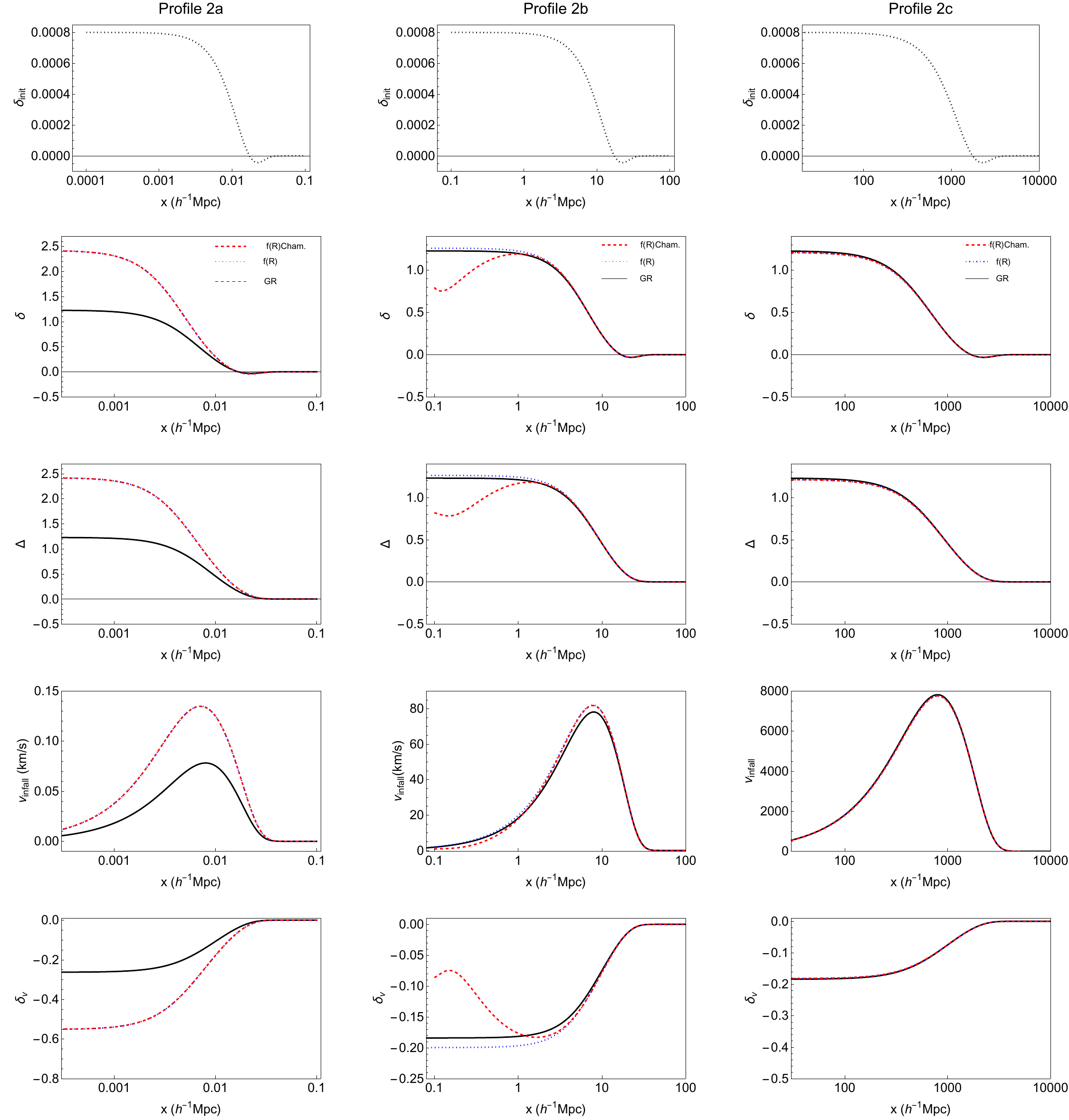}
 \caption{Non-linear evolution of density and velocity upto $a=1$.  The top panel plots the initial profiles at $a =0.001$. The second, third and fourth panels show the density contrast $\delta$, the spherically averaged density $\Delta$ and the infall velocity $v_{infall}$ respectively at $a=1$. The last panel shows the  fractional Hubble parameter for each shell $\delta_v$, which is a measure of the velocity divergence at that radial position. For profile 2a, there is an enhanced growth as compared to GR, but no effect of the Chameleon mechanism on the evolution. For profile 2c, the length scale is much larger than the Compton scale and the results are in agreement with GR. In this regime, the Chameleon screening is in effect only in those regions of high density gradient. Such steep gradients are absent in profile 2c. For profile 2b, the Chameleon mechanism suppresses growth significantly both compared to GR and compared to the case when Chameleon is not invoked. }
 \label{denvel}
 \end{figure}
 
  \begin{figure}
\includegraphics[width=16cm]{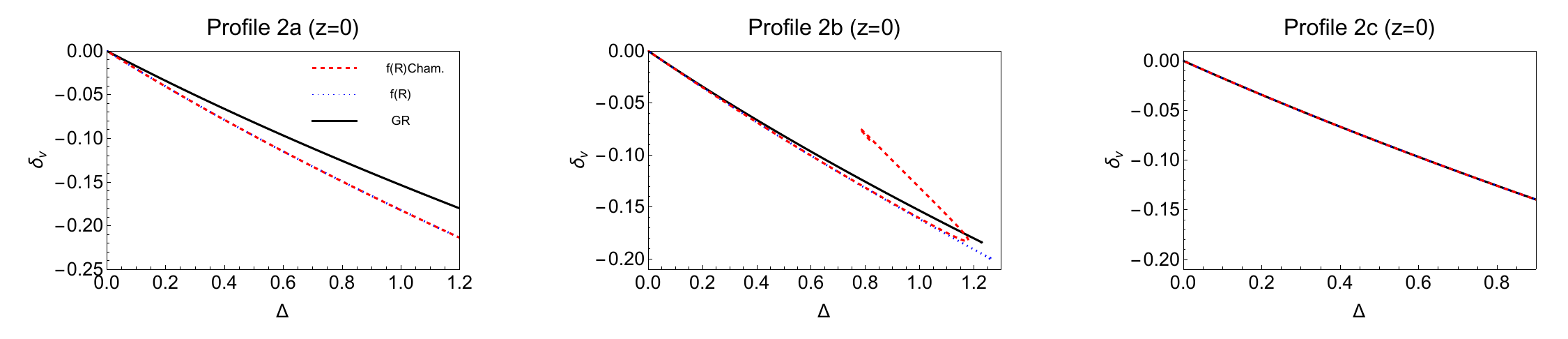}
 \caption{Evolution in the $\delta-\delta_v$ phasespace. As was seen in NC22, for the evolution of profile A and profile C there is a unique density-velocity divergence relation at late times. However, for scales where the Chameleon mechanism is active, even for a single profile, the $\delta-\delta_v$ curve does not stay monotonic, but becomes a multi-valued function. }
 \label{phasespace}
 \end{figure}

We consider smoothed, compensated top-hat profiles with three different scales \\  $x_{top} =  0.002,\, 2,\, 200 \, h^{-1}\mathrm{Mpc}$ (labels a,b,c resp.) and two values of smoothing parameters, $\sigma = 0.0025$ and $\sigma = 0.1$ (labels 1 and 2 resp.). The static iterative solution for $\chi$ is better illustrated with profiles 1a,b,c (figure 1) whereas for the dynamic evolution we present results with profiles 2a,b,c since physically realistic profiles are generally smooth. The details of the profiles are found in Appendix \ref{app:1}.

\subsection{Static perturbative solution for the $\chi$ field}
Figure \ref{Staticsoln} shows the solution for $\chi$ (scaled by $(H x_{top})^2$ to make it dimensionless) with and without including Chameleon screening at $a =1$ (fixed epoch). The solution with Chameleon is constructed by the series expansion \eqnref{chiexpansion}, kept upto the fourth order in $\epsilon$. The solution without Chameleon corresponds to the first term of the series: $\chi^{(1)}$. The top panel of the figure shows the initial density profile. The second and fourth  panels show the scaled $\chi$ and $\nabla \chi$  with and without screening. The third and fifth panel plot the fractional differences defined as 
\beq 
{\rm fractional}\;\; \chi = \frac{\chi^{Cham.} - {\chi^{no Cham.}}}{ |\chi^{Cham.}|_{,max}}  = \frac{\chi - {\chi^{(1)}}}{ |\chi|_{,max}}, 
\eeq
where, $|\chi|_{max}$ is the maximum absolute value of $\chi$ on the spatial grid in the presence of Chameleon. The fractional change in the force $\nabla \chi$ is defined similarly.
For profile 1a, $Q>>1$ for the entire perturbation and hence including the Chameleon screening makes no difference to the solution for $\chi$. For profile 1c, there is a clear difference in the solution for $\chi$ with and without the screening. Since the overdensity is positive inside the top-hat, the Compton scale is lower and hence $\chi$ which is proportional to ${\bar x}_C^2\delta$ is lower inside the perturbation when the screening mechanism is included. However, the force $\nabla_x \chi$, is the same in the inner regions whether or not Chameleon is included. It manifests itself only in presence of large density gradients and the fractional differences in $\nabla_x \chi$ show up only at the top-hat edge. For profile 1b, the perturbation is in the intermediate regime and sensitive to the Chameleon screening. This manifests as a small difference in the $\chi$ values at any given epoch, but its influence on the temporal evolution of the profile is seen in the next figure.

\subsection{Temporal evolution using the Eulerian-Lagrangian scheme}
Profiles 2a,b,c were evolved from $a=0.001$ to $a=1$ following the algorithm given in section \ref{sec:Method}. At the end of each step, the Ricci scalar was computed using \eqnref{eqnforR} and the comoving Compton wavelength ${\bar x}_C$ was computed using \eqnref{Comptondef2} and \eqnref{fofR}. We present results of these runs in this section. 
Figure \ref{Compton} examines the spatio-temporal evolution of the Compton scale and discusses its back-reaction on the density evolution. Figure \ref{denvel} presents the evolved density and velocity fields at $a=1$ and the same information is plotted in the density-velocity divergence phase-space in figure \ref{phasespace}.


\subsubsection{Spatio-temporal evolution of the Compton Scale}
Figure \ref{Compton} shows the initial density field (top-panel), the density evolved in time (second panel),  the comoving Compton scale ${\bar x}_C$ (third panel) and the $Q$ parameter 
(fourth panel) at four different epochs. In presence of perturbations, the Compton scale evolves both with time and space. 
We first note that the background value of the Compton scale increases with epoch at early epochs but slightly decreases near the epoch of dark energy dominance. This is clear from the asymptotic values (outer regions of the perturbation, where the density contrast is zero) of the brown, red, green and blue curves in the third and fourth panel. 

Consider profile 2a ($x_{top} = 0.002 h^{-1}$ Mpc). This profile is in the strong field regime throughout its evolution (as indicated by the value of $Q$). The Chameleon mechanism does not affect such a profile and the density evolution is the same with or without Chameleon. The Compton scale ${\bar x}_C$ does evolve with density; it is lower in the high density regions but this evolution does not have a back-reaction on the density field. The potential is 4/3 of its GR value and as we shall see in figure \ref{denvel}, the density grows faster than that in GR, as expected. 
Consider profile 2c ($x_{top} = 200 h^{-1}$ Mpc). The Compton scale is much smaller than the scale of the perturbation and this profile stays in the weak field regime throughout its evolution. The effect of the modification is suppressed by a factor of $Q^2$ with or without Chameleon and the evolution is very close to that in GR. In this regime, the potential $\chi \propto \delta$ and the effect of the modification will be experienced only where $\delta$ changes on a scale smaller than the Compton wavelength. This can be understood from figure \ref{Staticsoln} and was also  demonstrated in NC22 (fig. 12). 

Profile 2b has the most interesting evolution and is also the most astrophysically relevant situation.  At early epochs $Q <<1$ and the entire perturbation is in the weak field regime. The back-reaction of the spatially varying Compton scale does not affect the evolution and the shape of the perturbation is retained until $a \sim 0.3$. The evolution is also linear since $\delta$ is small. As evolution proceeds, $Q$ increases and the effect of the Chameleon mechanism starts to influence the evolution. At $a \sim 0.3$, ${\bar x}_C$ is lower inside the perturbation as compared to outside.  
A lower value of ${\bar x}_C$ in the inner shells means lower acceleration as compared to the outershells. Outer shells, as they approach the origin, encounter the relatively slower inner shells. This causes a mass build-up which is seen as a slight bump in the density field at $a \sim 0.5$. Consequently, the Chameleon scale also picks up a gradient in the inner regions: a relatively lower $\delta$ near the origin, implies a slight increase in ${\bar x}_C$ near the origin at $a \sim 0.5$. This in turn affects further evolution of the density. Due to the increased value of ${\bar x}_C$ near the origin, the 
 shells that are falling inward near the origin experience higher force and accelerate faster towards the origin followed by slightly slower shells causing a second density enhancement near the origin at $a \sim 0.75$ and the feature remains until $a\sim 1$  \footnote{ We evolved this profile beyond $a>1$ (not shown here) and found that the height of the peak increases with increase in final epoch. The enhancement near the origin did not propagate outwards and the final profiles looked qualitatively similar to the one shown here. Whether or not this changes with spatial resolution remains to be checked.}.
 The density enhancement at the edge of the top-hat is a known feature of the Chameleon mechanism and has been observed in earlier investigations \cite{borisov_spherical_2012,kopp_spherical_2013}. However, to the best of our knowledge, the enhancement in the interior of the perturbation has not been reported in the literature so far. 


\subsubsection{Non-linear density and velocity fields}
Figure \ref{denvel} shows the density and velocity profiles at $a=1$ for the same three initial conditions for three cases (i) $f(R)$ with Chameleon, (ii) $f(R)$ without Chameleon and (iii) standard GR ($\Lambda$CDM). The overdensity ($\delta$) and the spherically averaged overdensity ($\Delta$, defined in \eqnref{Deltadef}) 
are in the second and third panels respectively. As expected, for profile 2a, the density grows faster than that in GR, whereas for profile 2c, it is indistinguishable from the GR evolution since the Compton scale is very small compared to the scale of the perturbation. The evolution of the density for profile 2b has been discussed in detail in the earlier section. It is useful to note that while the profile without the Chameleon predicts slightly higher growth than the GR evolution, the Chameleon mechanism suppresses growth everywhere compared to GR. The fourth panel shows the infall velocity defined as the negative of the peculiar velocity
\beq
v_{infall}(q, a)  = - v(q, a). 
\eeq
For a perfect top-hat, the expected infall velocity is usually linear in the inner shells; but for the smooth profiles considered the peculiar velocity is maximum near the edge of the top-hat and then decreases monotonically in all three cases. 
For profile 2b, the difference in the evolution is not well-captured by the infall velocity. Instead, it is instructive to compute the profiles of the fractional Hubble parameter of the shell defined as 
\beq 
\delta_v=\frac{1}{H} \frac{\dot r}{r} - 1 = \frac{1}{H} \frac{v(q, a)}{a x(q, a)} = \frac{A'}{A}. 
\eeq
For a constant density profile, this is related to the scaled velocity divergence as $\Theta = 3 \delta_v$. The profile for $\delta_v$ shows a clear distinction between the evolution with and without the Chameleon mechanism. 


\subsubsection{Phase-space evolution}
In figure \ref{phasespace}, we plot the $(\Delta, \delta_v)$ pairs at $z=0$ or $a=1$. In standard GR, in linear theory, the density-velocity divergence relation (DVDR) is linear:  $\Theta = - a H f(\Omega_m) \Delta$. There have been extensions of this relation in the non-linear regime, starting with the early investigations based on perturbation theory (for e.g., \cite{bernardeau_quasi-gaussian_1992, chodorowski_large-scale_1997,chodorowski_weakly_1997,susperregi_cosmic_1997}) as well a numerical simulations (for e.g., \cite{bernardeau_new_1996, chodorowski_recovery_1998, zaroubi_wiener_1999,kudlicki_reconstructing_2000,ciecielag_gaussianity_2003,kitaura_estimating_2012}). 
This was later extended using spherical collapse by \cite{bilicki_velocity-density_2008}, where they relied on the exact analytic solutions of the spherical top-hat model. In 2013, \cite{nadkarni-ghosh_non-linear_2013}  investigated this relation also using spherical collapse, but not by relying on the closed form solutions of spherical geometry. 

Instead by imposing the condition of `no perturbations at the bang time' one can compute $(\delta, \delta_v)$ pairs which trace out a unique curve in the two-dimensional $\delta-\delta_v$ phase space. By examining the underlying dynamical system, it was shown that this curve (called the Zeldovich curve) is an invariant of the dynamics. This means that no matter what the initial conditions may be, the late-time non-linear density and velocity perturbations will always lie on this curve indicating that this is the late-time non-linear density velocity divergence relation (DVDR), the correct non-linear extension of the linear theory relation in spherical symmetry. The fitting form for this curve provided by \cite{nadkarni-ghosh_non-linear_2013}, which was based on earlier forms provided by \cite{bernardeau_quasi-gaussian_1992} and \cite{bilicki_velocity-density_2008} was also found to hold true for early dark energy models with a varying equation of state \citep{mandal_one-point_2020}. 
In NC22, this dynamics was investigated for $f(R)$ models without the Chameleon mechanism. In the strong field regime, where the equation has the same structure as GR a unique non-linear DVDR relation exists. In the weak field regime, the corrections to the dynamics are negligible and the GR relation is reproduced. But, in general, in the intermediate regimes, the structure of the equations is such that the relation is dependent upon the shape of the initial profile, violating the invariant property seen in standard gravity. However, the relation always remains one-to-one. In the presence of the Chameleon mechanism, for profile 2b, the relation is no longer one-to-one even for a single profile. A given $\Delta$ value can have multiple $\delta_v$ values depending upon the location in the perturbation. This is fundamentally related to the non-linear feedback in built in the Chameleon mechanism. The density decides the value of the Compton scale, which in turn influences the density evolution. 
This behaviour is different from the scatter in the DVDR relation observed in standard gravity in triaxial systems, where the multi-valued behaviour originates because of the three-dimensional nature of the dynamics \citep{nadkarni-ghosh_phase_2016}. 


\section{Discussion and Conclusion} 
\label{sec:Conclusion}
In this paper, we investigate the effect of the Chameleon mechanism on smooth, compensated, top-hat density profiles evolving in a Universe with $f(R)$ gravity. In particular, we focus on the Hu-Sawicki model with parameters which are compatible with solar system constraints. This work builds on earlier investigation (NC22) by one of the authors which considered the same system but ignored the Chameleon screening. The new addition to the earlier investigation is a perturbative scheme for the scalar field potential $\chi = \Phi - \Psi$, which satisfies a non-linear differential equation. Chameleon screening effects in spherical collapse models have been discussed in the past, but the structure of the equations necessitates sophisticated two-point boundary value ODE solvers \cite{borisov_spherical_2012,kopp_spherical_2013}. The advantage of our method is that, although approximate, it is computationally inexpensive since each iteration uses an analytic solution to the linear differential operator presented in \cite{Nadkarni_Ghosh_2022} (NC22). 
 In spherical symmetry we used the analytic forms which were developed in NC22, but the method is easily generalizable to generic periodic (random) 3D initial conditions. In that case, in each successive approximation for $\chi$ involves solving a Poisson-like equation which can be easily done using FFTs. Such a scheme can be coupled to other numerical methods to solve for the dynamical equations - either 
Lagrangian Perturbation Theory  \citep{nadkarni-ghosh_extending_2011,nadkarni-ghosh_modelling_2013} or even N-body simulations which move the particles having computed the force-fields. We leave this for future investigation. 

The dynamics of the perturbations depend sensitively on the parameter $Q$, defined to be the ratio of the Compton scale to the top-hat scale. In the presence of Chameleon screening, 
$Q$ varies both in space and in time. When $Q>>1$ (referred to as the strong field limit) the equation for $\chi$ is insensitive to the  Chameleon screening and the dynamics is equivalent to standard GR with a Newton's constant $G$ enhanced by a factor of 4/3. When $Q<<1$ (weak field) the effects of the modification are felt only if density changes on scales of the order of the Compton scale. Given that the Compton scale is small in this regime, the modification will manifest itself only if the density field has sharp gradients. 
When the scale of the perturbation matches the Compton scale, the effect of the modification is maximally visible since the evolving perturbation changes its shape considerably. The Compton scale drops inside the perturbation giving rise to a lower force and inside the perturbation the growth is in fact suppressed as compared to GR. 
Incoming mass shells accumulate near the edge of the top-hat giving rise to a peak near the edge. In this paper, we illustrate these three regimes with three values of the top-hat scale 
 $x_{top} = 0.002, 2, 200 h^{-1}$Mpc. The smallest scale $x_{top} = 0.002$ stays in the strong field regime throughout the evolution giving a higher density contrast compared to GR as expected. The largest scale $x_{top} = 200 h^{-1}$ Mpc stays in the weak-field regime throughout and is almost indistinguishable from GR. The profile with $x_{top} = 2 h^{-1}$ Mpc, shows the density enhancement near the edge as expected. \cite{borisov_spherical_2012} and \cite{kopp_spherical_2013} also report similar findings with \cite{borisov_spherical_2012} showing suppressed growth inside the top-hat compared to GR. However, in our investigation, we observe a small rise in density near the centre of the smooth top-hat. The gradient in density inside the perturbation in turn induces a gradient in the Compton scale - lower density contrasts near the origin give rise to a higher Compton scale which in turn gives rise to a bump near the origin. We are able to explain the origin of the feature and do not attribute it to numerics. To the best of our knowledge, this feature has not been reported in the literature so far; perhaps because most efforts involving spherical collapse focus on calculating $\delta_c$ and not always on modelling the entire density profile. An in-depth investigation of such a feature for different values of initial amplitudes and profile shapes is worth undertaking. 
 
 Finally, we investigate the dynamics of the perturbations in the two-dimensional density-velocity divergence phase space. Unlike the case for GR or $f(R)$ without Chameleon, in the screening model the same density can correspond to two different values of velocity divergence depending upon the location in the profile. This is reminiscent of a hysteresis loop (albeit open in this case) which can arise from the non-linear feedback built into the Chameleon screening. The density field affects the Compton scale which in turn affects the density growth. 
Understanding how, if at all, such `hysteretic' signatures exist and whether they manifest real data sets would be an interesting future project. In conclusion, Chameleon screening provides a rich structure in the dynamics of cosmological perturbations inducing various new features compared to the standard GR evolution. While our investigation relies on spherically symmetric systems, it is conceivable that some of these features are generic and observable in N-body simulations and we hope that the insights gained from this special case can help analyse the results of simulations in novel ways. 

 \appendix
 \renewcommand{\thefigure}{A\arabic{figure}}
\setcounter{figure}{0}
 
\section{Numerical profiles and convergence tests}
\subsection{The compensated top-hat}
 \label{app:0}
The compensated top-hat is a one-dimensional function represented as
\bea 
\delta_{top}(x) &=& A\;\;\; \;\;\; \;\;\;\; 0< x \leq x_{top}\\
&=& -1 \;\;\;  x_{top} < x \leq x_u \\
&=& 0 \;\;\; {\rm otherwise,}
\eea
where $A$ is the amplitude of the top-hat. $x_{top}$ and $x_u$ are the boundaries of the overdense region and the underdense compensating region and are related to through conservation of mass as 
\beq
(1+A) x_{top}^3 = x_u^3. 
\eeq
A smooth top-hat is obtained by the following function 
\beq 
\delta_{smooth}(x) = \frac{1}{\sqrt{2 \pi \sigma^2 x^2}} \left\{\int_0^{x_{top}} A \left(e^{-\frac{(x-y)}{2 \sigma^2}} - e^{-\frac{(x+y)}{2 \sigma^2}}\right) y dy + \int_{x_{top}}^{x_u} (-1) \left(e^{-\frac{(x-y)}{2 \sigma^2}} - e^{-\frac{(x+y)}{2 \sigma^2}}\right) y dy \right\}, 
\eeq
where $\sigma$ is the smoothing parameter. In this paper we have used three different scales  $x_{top} =  0.002, 2, 200,h^{-1},\mathrm{Mpc}$ with two values of smoothing parameter $\sigma = 0.1$ and $\sigma = 0.0025$.  We refer to the profiles as 1(2) a,b,c depending upon the value of $x_{top}$ and the smoothing parameter (see table \ref{table1}).

\subsection{Convergence of the iterative solution for $\chi$. } 
\label{app:1}

 \begin{figure}
 \includegraphics[width=16cm]{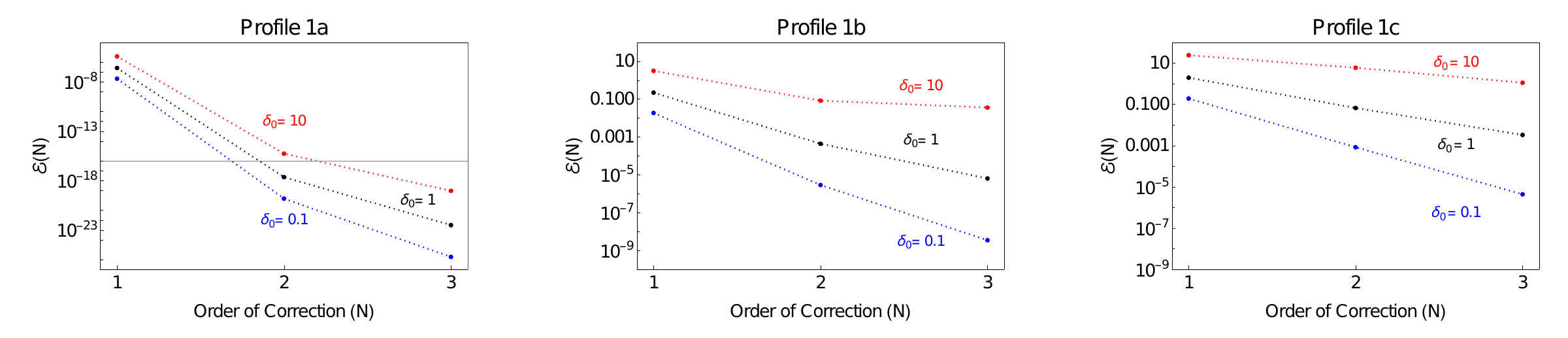} \\
 \includegraphics[width=16cm]{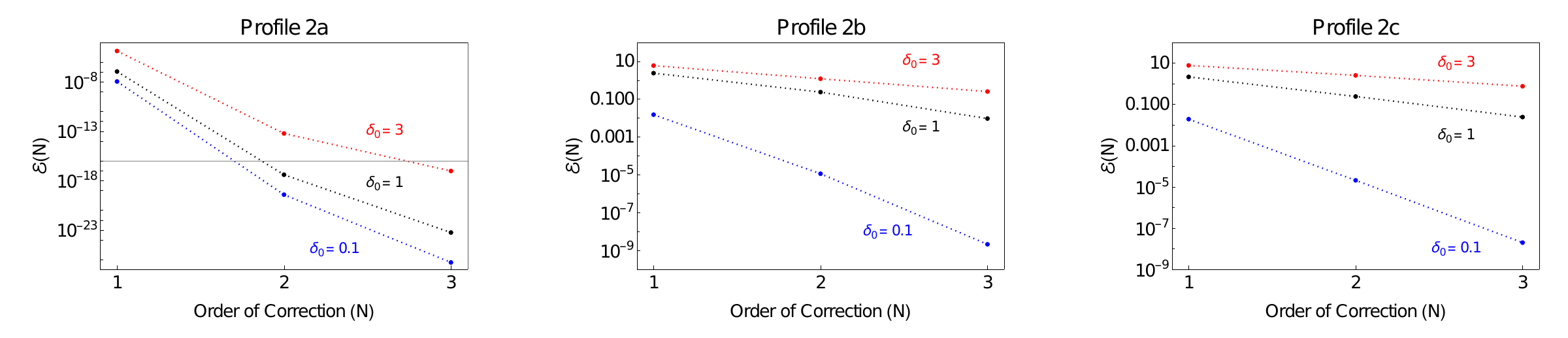}
 \caption{Convergence test of the Taylor expansion. Comparison of successive terms in the Taylor expansion of $\chi$ at a=1 using 500 spatial points. $\mathcal{E}(N)$ has been calculated up to $\delta_0 =30$ and we observe convergence up to $\delta_0 =15$.  }
 \label{orderC}
 \end{figure}
 
 We tested the perturbative solution for $\chi$ presented in section \ref{Chisoln}, for  all six profiles upto the fourth order solution for  $\chi$. Since we are solving the spatial equation, without loss of generality, we set $a=1$. Given the form of $\chi$ in \eqnref{chiexpansion}, the Cauchy error between successive approximations for $\chi$ is given as 
 \beq
 \mathcal{E}(n) = \text{Max}\{ \chi^{(n+1)}(x)\}, 
\eeq
where $\chi^{(2,3,4)}$ are the second, third and fourth order corrections to the first order solution $\chi^{(1)}$, which is the solution in the absence of Chameleon  screening. 
 The maximum is taken over all radial shell positions $x$. \capfigref{orderC} plots the error $ \mathcal{E}(n)$ vs $n$ for three values of the top-hat amplitude for each of the six profiles. 
 The dashed line indicates the level of numerical precision. It is clear that profiles in the strong field regime (1a and 2a) are not affected much by the screening and the errors remain low falling below the numerical precision for higher orders. 
 For profiles 1b,c and 2b,c, the errors decrease with order of expansion indicating convergence.  For a larger value of initial $\delta$, the convergence is slower, which is expected since the magnitude of the correction terms is directly proportional to $\delta$. However, for values of initial $\delta$ greater than a certain $\delta_0$, convergence was not observed. This value was 
 $\delta_0 \sim 15$ for $\sigma = 0.1$ and $\delta_0 \sim 5$ for $\sigma = 0.0025$.
 

\subsection{Convergence of the algorithm for the temporal evolution }
\label{app:2}
\begin{figure}
 \includegraphics[width=16cm]{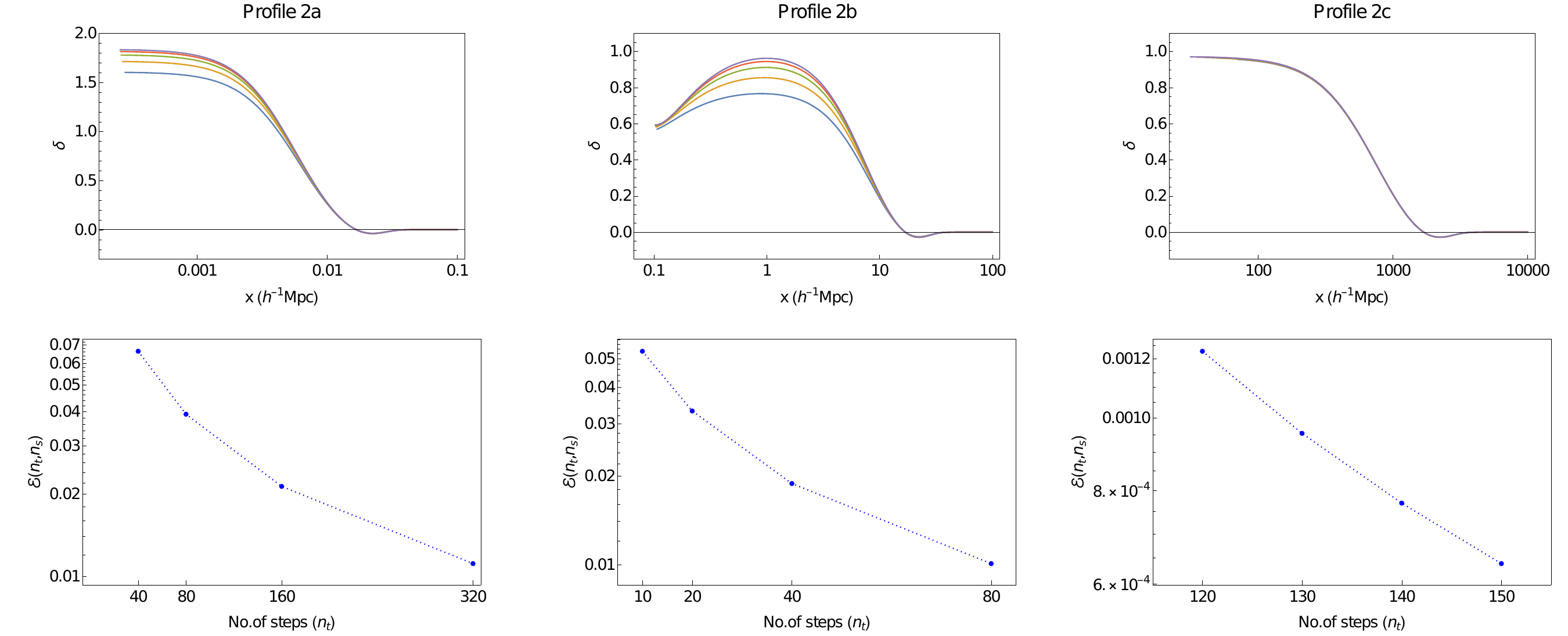}
 \caption{Convergence test of the full algorithm. $\mathcal{E}(N)$ is plotted against $n_t$. The grid sizes corresponding to the profiles are provided in Table \ref{tab:B1}.}
 \label{convwithNt}
 \end{figure}

In NC22, we performed convergence tests of the temporal evolution algorithm which ignored the Chameleon mechanism. The same parameter values for the number of time steps ($N_t$) and the number of radial shells ($N_s$) are not expected to work when we include the Chameleon except for profiles 1a and 2a which are almost insensitive to the screening. 
We considered only the profiles 2a,b,c since we present results only based on this profile. Realistic profiles are far from top-hats. 
We chose the parameters, primarily on a trial-and-error basis. They can be estimated from imposing the Courant–Friedrichs–Lewy (CFL) condition for numerical stability \citep{NumRec}. This reads 
\beq
\frac{u \Delta t}{\Delta x }  \leq 1
\eeq
where $u$ is the typical velocity, $\Delta t $ and $\Delta x$ are the temporal and spatial resolution respectively. In our case, $\Delta t \equiv \Delta \ln a$ and $\Delta x \equiv \Delta \ln x/(1 Mpc)$. However, in our case the velocity is a spatially evolving quantity and a rigorous application of this condition would imply an adaptive-time step scheme, which further complicates implementation. Hence we instead chose a trial-and-error path and checked for stability at each time step. We have checked that the time step criterion satisfies 
\beq
\frac{u_{max} \Delta \ln a}{x_{top} H_0\Delta \ln x} \sim 1. 
\eeq
where $u_{max}$ is the maximum infall velocity at $a=1$, $x_{top}$ is the scale of the perturbation and $\Delta \ln a$ and $\Delta \ln x$ are the resolution in $\ln a$ and $\ln x$ spaces respectively. 
We found that for profile 2c, numerical errors destroyed continuity of the solution for $\chi$ in the initial stages of evolution. To remedy this, we relied on the approximate answer. It can be seen that in the weak-field limit ($Q<<1$), $\chi^{(n)} =  {\bar x}_C^2 S^{(n)}$, where $S^{(n)}$ is the form of the source term for the $n$-th approximation. Numerically, we use this solution for $\chi$ for the initial steps until $Q<0.027$. After this value of $Q$, the numerical solution seems to stabilize and we use the analytic answer which solves the full equation for $\chi$. Both $N_s$ and $N_t$ were simultaneously increased according to the values shown in the table. While increasing $N_t$ for a fixed $N_s$, ensures numerical stability, we found that it also compromises the resolution in the inner parts of the top-hat. \capfigref{convwithNt} shows the Cauchy error defined as the difference between successive approximations as a function of $N_t$ for profiles 2abc. The error decreases in all cases indicating that the algorithm converges. The plots shown in figures \ref{Compton} and \ref{denvel} use the final values for $N_s$ and $N_t$ given in table \ref{tab:B1}, for e.g., profile 2a was evolved with $N_s = 1600$ and $N_t = 320$. 


\begin{table}
\centering
\begin{tabular}{|c |c |c |c |c |c|}
\hline
Profile Name & $\sigma$ & $x_{\mathrm{top}}$ &
$L_{\min} \; (h^{-1}\,\mathrm{Mpc})$ &
$L_{\max} \; (h^{-1}\,\mathrm{Mpc})$ \\
\hline
1a & 0.0025 & 0.002 & 0.0001 & 0.1 \\
1b & 0.0025 & 2     &  0.1    & 100 \\
1c & 0.0025 & 200   & 10     & 10000 \\
\hline
2a & 0.1    & 0.002 & 0.0001 & 0.1 \\
2b & 0.1    & 2     & 0.1    & 100 \\
2c & 0.1    & 200   &  10     & 10000 \\
\hline
\end{tabular}
\caption{Table denoting the parameters of the two profiles used. In the numerical implementation, $L_{min}$ and $L_{max}$ denote the minimum and maximum values of the grid points along the radial direction. }
\label{table1}
\end{table}

 \begin{table}
\centering
\begin{tabular}{| c| c| c| c| c| c|}
\hline
Profile used & $A$ & $N_s$ & $N_t$ &
$a_{\mathrm{switch}}$ & $a_{\mathrm{final}}$ \\
\hline
 2a      & 0.0008    & \{100, 200, 400, 800, 1600\}    & \{20, 40, 80, 160, 320\}      & 0.1   & 1 \\
 2b      & 0.0008    & \{200, 400, 800, 1600,3200\}    & \{5, 10, 20, 40, 80\}      & 0.1   & 1 \\
 2c      & 0.0008    & \{300, 400, 500, 600,700\}    & \{110, 120, 130, 140, 150\}      & 0.1   & 1 \\
\hline
\end{tabular}
\caption{Table listing the profiles and numerical parameters used for testing the convergence of the algorithm. Refer to \figref{convwithNt}.}
\label{tab:B1}
\end{table}

\begin{table}
\centering
\begin{tabular}{|c| c| c| c| c| c| c|}
\hline
Figure & Profile used & $A$ & $N_s$ & $N_t$ &
$a_{\mathrm{switch}}$ & $a_{\mathrm{final}}$ \\
\hline
figure 1   & 1a,b,c      & 7                       & 3000    & --      & --   & -- \\
figure 2, 3, 4  & 2a,b,c      & 0.0008             & 1600,3200,700    & 320,80,150      & 0.1   & 1  \\
\hline
\end{tabular}
\caption{Table listing the profiles and numerical parameters used to generate figures in the text.}
\label{tab:A2}
\end{table}

\acknowledgments
S. N-G would like to thank IIT Kanpur for the Initiation Grant (SSA/2022232) and the Department of Science and Technology, ANRF (previously Science and Engineering Research Board) for the MATRICS grant (No.MTR/2023/001376 ). T.R.V. would like to thank the SURGE programme at IIT Kanpur, which enabled the start of this project.

\newpage

 \bibliographystyle{JHEP}
 \bibliography{modgrav,book}
\newpage

\end{document}